\documentclass[conference]{IEEEtran}
\ifCLASSINFOpdf
\else
\fi
\usepackage{algorithm}
\usepackage{float}
\usepackage{algorithm,algpseudocode}
\usepackage{graphicx}
\usepackage{array}
\usepackage{float}
\usepackage{lipsum}
\usepackage{authblk}
\usepackage{mathtools}
\usepackage{authblk}

\usepackage{subcaption}

\begin{document}
%
\title{NeuTM: A Neural Network-based Framework for Traffic Matrix Prediction 
in SDN
}
%
%
%

\author{Abdelhadi Azzouni\thanks{abdelhadi.azzouni@lip6.fr}}

\author{Guy Pujolle\thanks{guy.pujolle@lip6.fr}}

\affil{LIP6 / UPMC; Paris, France  \{abdelhadi.azzouni,guy.pujolle\}@lip6.fr}

\maketitle

\begin{abstract}
This paper presents NeuTM, a framework for network Traffic Matrix (TM) prediction based on 
Long Short-Term Memory Recurrent Neural Networks (LSTM RNNs). 
TM prediction is defined as the problem of estimating 
future network traffic matrix from the previous and achieved network
traffic data. It is widely used in network planning, resource management and network security.
Long Short-Term Memory (LSTM) is a specific recurrent neural network (RNN) architecture  
that is well-suited to learn from data and classify or predict time series 
with time lags of unknown size. LSTMs have been shown to model 
long-range dependencies more accurately than conventional RNNs.
NeuTM is a LSTM RNN-based framework for 
predicting 
TM in large networks. 
By validating our framework on real-world data from G\'EANT network, 
we show that our model converges quickly and gives state of the
art TM prediction performance.


\end{abstract}


 {\bf { \it keywords - }}
Traffic Matrix, Prediction, Neural Networks, Long Short-Term Mermory, Software Defined Networking

%
\IEEEpeerreviewmaketitle

\section{Introduction}


Having an accurate and timely network TM is essential for most network 
operation/management tasks such as traffic accounting, short-time traffic scheduling or re-routing, 
long-term capacity planning, network design, and network anomaly detection.  
For example, to detect DDoS attacks in their early stage, it
is necessary to be able to detect high-volume traffic clusters in
real-time, which is not possible relying only on current monitoring tools. 
Another example is, upon congestion occurrence in the network, 
traditional routing protocols cannot react immediately to 
adjust traffic distribution, resulting
in high delay, packet loss and jitter. 
Thanks to the early warnings, a proactive prediction-based approach would be faster, in terms
of high-volume traffic detection and DDoS prevention. Similarly, 
predicting network congestion is more effective than reactive
methods that detect congestion through measurements, only
after it has significantly influenced the network operation.

 
Network traffic is characterized by: self-similarity, multiscalarity, 
long-range dependence and a highly nonlinear nature (insufficiently modeled by Poisson and Gaussian models for example).
These statistical characteristics determine the traffic's predictability \cite{selfsimilar}.

Several methods have been proposed for
network traffic prediction and can be classified into two
categories: linear prediction and nonlinear prediction. 
The ARMA/ARIMA model \cite{forecasting}, \cite{studypredict}, \cite{mpegvid}  and the
Holt–Winters algorithm \cite{forecasting} are the most
widely used traditional linear prediction methods. 
Nonlinear
forecasting methods commonly involve neural networks (NN) \cite{forecasting}, \cite{mpeg2},
\cite{mpeg3}. The experimental results from \cite{evaluation}
show that nonlinear traffic prediction based on NNs 
outperforms linear forecasting models (e.g. ARMA, ARAR, HW).
\cite{evaluation} suggests that if we take into
account both precision and complexity, the best results are
obtained by a Feed Forward Neural Network predictor with multiresolution learning
approach. However, most of the research using neural networks for network traffic prediction
aims to predict the aggregate traffic value. In this work, our goal is to predict 
the traffic matrix which is a far more challenging task.

 

%

Unlike feed forward neural networks (FFNN), 
Recurrent Neural Network
(RNNs) have cyclic connections over time. 
The activations from each time step are stored in the internal
state of the network to provide a temporal memory.
This capability makes RNNs better
suited for sequence modeling tasks such as time series prediction and
sequence labeling tasks. 
Particularly, Long Short-Term Memory (LSTM) is a powerful RNN architecture that was recently designed by 
Hochreiter and Schmidhuber \cite{lstmoriginalpaper} to address the  
vanishing and exploding gradient problems \cite{vanishing} that conventional RNNs suffer from.
RNNs (including LSTMs) have been successfully used for handwriting recognition \cite{handwritinglstm}, 
language modeling, phonetic labeling of acoustic frames \cite{sakgoogle}.


Our contribution in this paper is threefold. 
\begin{itemize}
 \item First,
we present, for the first time, a LSTM based framework for large scale 
TM prediction. 
 \item Second, we implement our framework and deploy it on a Software Defined Network (SDN) 
and train it on real world data using G\'EANT data set.
 \item Finally, we evaluate our LSTM models at different configurations.
We also compare our model to traditional models and show that LSTM models converge quickly and give state of the
art TM prediction performance. 




\end{itemize}

Note that we do not address the problem of TM estimation 
in this paper and we suppose that historical TM data is already accurately obtained.

The remainder of this paper is organized as follows: Section \ref{timeseriesprediction} summarizes 
time-series prediction techniques. 
LSTM architecture and equations are detailed in section \ref{lstm}.
The process of feeding the LSTM model and predicting TM is described in section \ref{tmpredictionusinglstm}.
Evaluation and results are presented in section \ref{experiments}. Related work is discussed in section \ref{relatedwork}
and the paper is concluded by section \ref{conclusion}.

\section{Time Series Prediction}\label{timeseriesprediction}
For completeness sake, we give a brief summary of various linear 
predictors based on traditional statistical techniques. We use the same 
notation and definitions as in \cite{evaluation} and we refer to the original paper and to
\cite{introductiontimeseries} for a thorough background.  
Then we discuss NNs usage for time series prediction.	 

\subsubsection{Linear Prediction}
\paragraph{ARMA model}

The time series $\{X_t\}$ is called an ARMA(p, q) process if
$\{X_t\}$ is stationary and 
\begin{equation} \label{delta}
       X_t -  \phi _1 X_{t-1}-. . .- \phi_p X_{t-p} = Z_t +  \theta  _1 Z_{t-1} +. . .+  \theta  _q Z_{t-q}
  \end{equation}
where $\{Z _t\} \approx  WN(0,  \sigma ^2 )$ is white noise with zero mean and
variance $\sigma^2$ and the polynomials  $\phi (z) = 1 -  \phi_1 z - . . . -  \phi_p z^p$
and $\theta (z) = 1 +  \theta_1 z + . . . +  \theta_q z^q$ have no common factors.
Predictions can be made recursively using:
$ \widehat{X}_{n+1}= 
 \begin{cases} 
\sum_{j=1}^n   \theta _{nj} (X_{n+1-j}- \widehat{X}_{n+1-j})  &  if  1 \leq n \leq m) \\
\sum_{j=1}^q   \theta _{nj} (X_{n+1-j}- \widehat{X}_{n+1-j})\\
 +  \phi _1 X_n+..+ \phi _p X_{n+1-p} & if  n \geq m
\end{cases} 
$
where $m = max(p, q)$ 
and $\theta_{nj}$ is determined using the
innovations algorithm.

\paragraph{ARAR algorithm}
The ARAR algorithm applies memory-shortening transformations,
followed by modeling the dataset as an AR(p)
process: 
$X_t = \phi_1 X_{t-1} + ..+ \phi_p X_{t-p} + Z_t$.
The time series $\{Y_t\}$ of long-memory or moderately long-
memory is processed until the transformed series can be
declared to be short-memory and stationary:
\begin{equation} \label{phi}
 S_t = \psi(B)Y_t = Y_t + \psi_1 Y_{t-1} + . . . + \psi_k Y_{t-k}
 \end{equation}
The autoregressive model fitted to the mean-corrected series
$X_t = S_t - \overline{S} ̄$, $t =\overline {k + 1, n}$, where $\overline{S}$ 
represents the sample mean for $S_{k+1} , . . . , S_n$ , is given by $\phi(B)X_t = Z_t$ ,
where $\phi (B) = 1- \phi_1 B -  \phi_{l_1} B^{l_1} -  \phi_{l_2} B^{l_2} -  \phi_{l_3} B^{l_3}, \{Z _t\} \approx  WN(0,  \sigma ^2 )$, 
while the coefficients $\phi_j$ and the variance $\sigma^2$ are
calculated using the Yule–Walker equations described in \cite{introductiontimeseries}.
We obtain the relationship:
\begin{equation} \label{phi}
 \xi (B)Y_t =  \phi (1)  \overline{S}  + Z_t 
 \end{equation}
where  $\xi (B)Y_t =  \psi (B) \varphi (B) = 1 +  \xi_1B + . . . + \xi_{k+l_3} B^{k+l_3}$
From the following recursion relation we can determine the
linear predictors
\begin{equation} \label{phi}
   P_n Y_{n+h} = - \sum_{j=1}^{k+l_3} \xi P_n Y_{n+h-j} + \phi(1) \overline{S}  \quad h\geq1
 \end{equation}
with the initial condition $P_n Y_{n+h} = Y_{n+h}$ for $h \leq 0$. \\

\paragraph{Holt–Winters algorithm}
The Holt–Winters forecasting algorithm is an exponential
smoothing method that uses recursions to predict the
future value of series containing a trend. 
If the time series has a trend, then the forecast function is:
\begin{equation} \label{phi}
    \widehat{Y}_{n+h} = P_n Y_{n+h} =  \widehat{a}_n +  \widehat{b}_n h 
 \end{equation}
where $\widehat{a}_n$ and $\widehat{b}_n$ are the estimates of the level of the trend
function and the slope respectively. These are calculated using
the following recursive equations:
\begin{equation} \label{phi}
\begin{cases}
 \widehat{a}_{n+1} =  \alpha Y_{n+1} + (1 -\alpha)( \widehat{a}_n +  \widehat{b}_n) \\
 \widehat{b}_{n+1} =  \beta ( \widehat{a}_{n+1}-  \widehat{a}_n ) + (1 -  \beta )  \widehat{b}_n 
\end{cases} 
\end{equation}
Where $ \widehat{Y}_{n+1} = P_n Y_{n+1} =  \widehat{a}_n +  \widehat{b}_n $ 
represents the one-step
forecast. The initial conditions are:  $\widehat{a}_2 = Y_2$ and 
$\widehat{b}_2 = Y_2 - Y_1$. 
The smoothing parameters $\alpha$ and $\beta$ can be chosen
either randomly (between 0 and 1), or by minimizing the sum
of squared one-step errors  $\sum_{i=3}^n  (Y_i - P_{i-1} Y_i )^2$ \cite{introductiontimeseries}.

\subsubsection{Neural Networks for Time Series Prediction} 
Thanks to their strong self-learning and their ability to learn
complex non-linear patterns,
Neural Networks (NNs) are widely used for
modeling and predicting time-series.  
NNs are capable of estimating almost any linear or non-linear 
function in an efficient and stable manner, when the underlying
data relationships very complex.
Unlike the techniques presented above,
NNs rely on the
observed data rather than on an analytical model.
%
Furthermore, The architecture and the parameters of a NN are determined
solely by the dataset. 

A neural network consists of interconnected nodes, called
neurons. The interconnections are weighted and the weights are also called parameters.
Neurons are organized in layers: a) an input layer,
b) one or more hidden layers and c) an output layer. 
The most popular NN architecture is feed-forward in which the
information goes through the network only in the forward
direction, i.e. from the input layer towards the output layer, as
illustrated in figure \ref{ffnnn}.

\section{Long Short Term Memory Neural Networks} \label{lstm}

\begin{figure*}[!htb]
\minipage{0.32\textwidth}
  \includegraphics[scale=0.3]{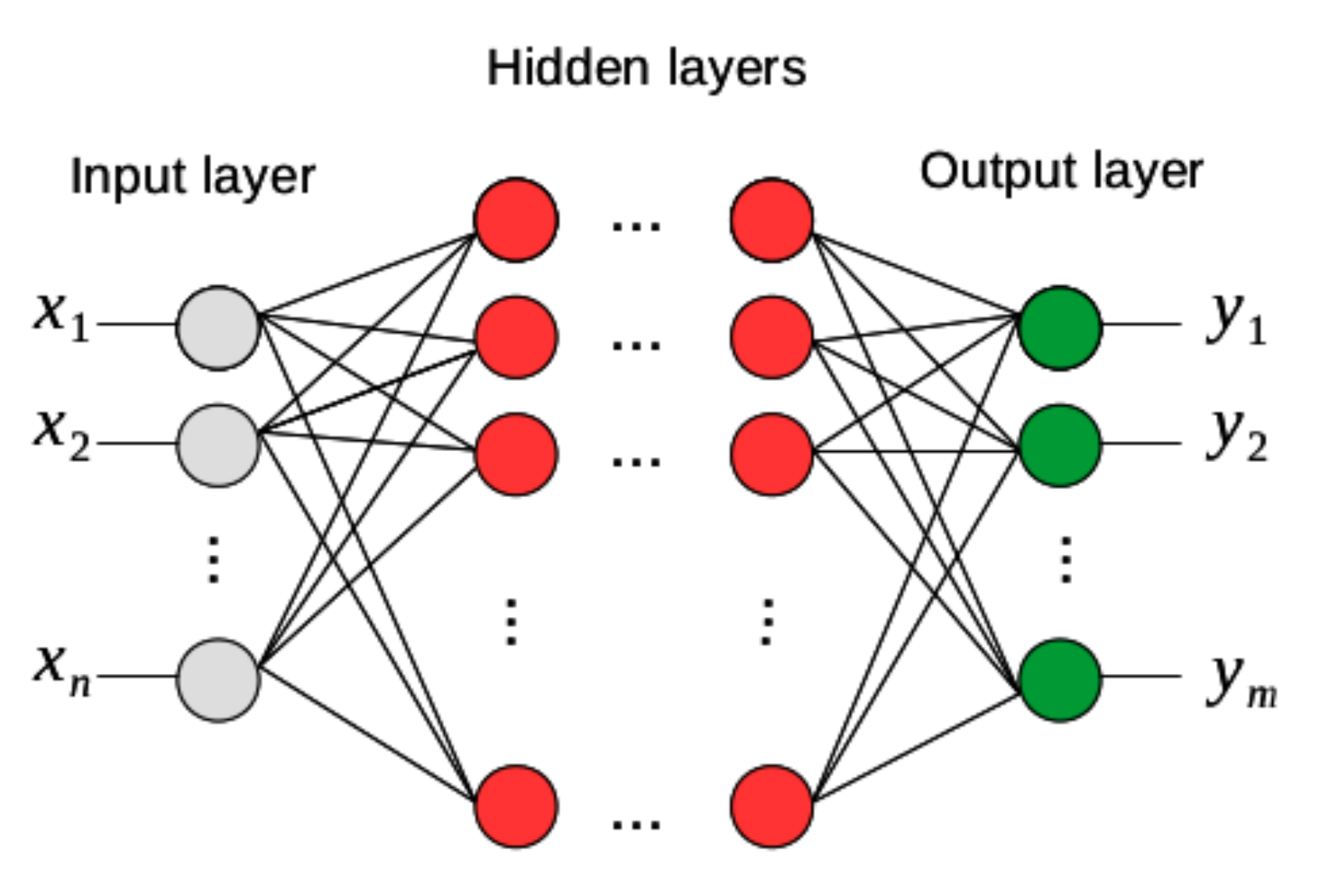}
  \caption{\label{fig:lldppacket} Feed Forward Deep Neural Network}\label{ffnnn}
\endminipage\hfill
\minipage{0.32\textwidth}
  \includegraphics[scale=0.3]{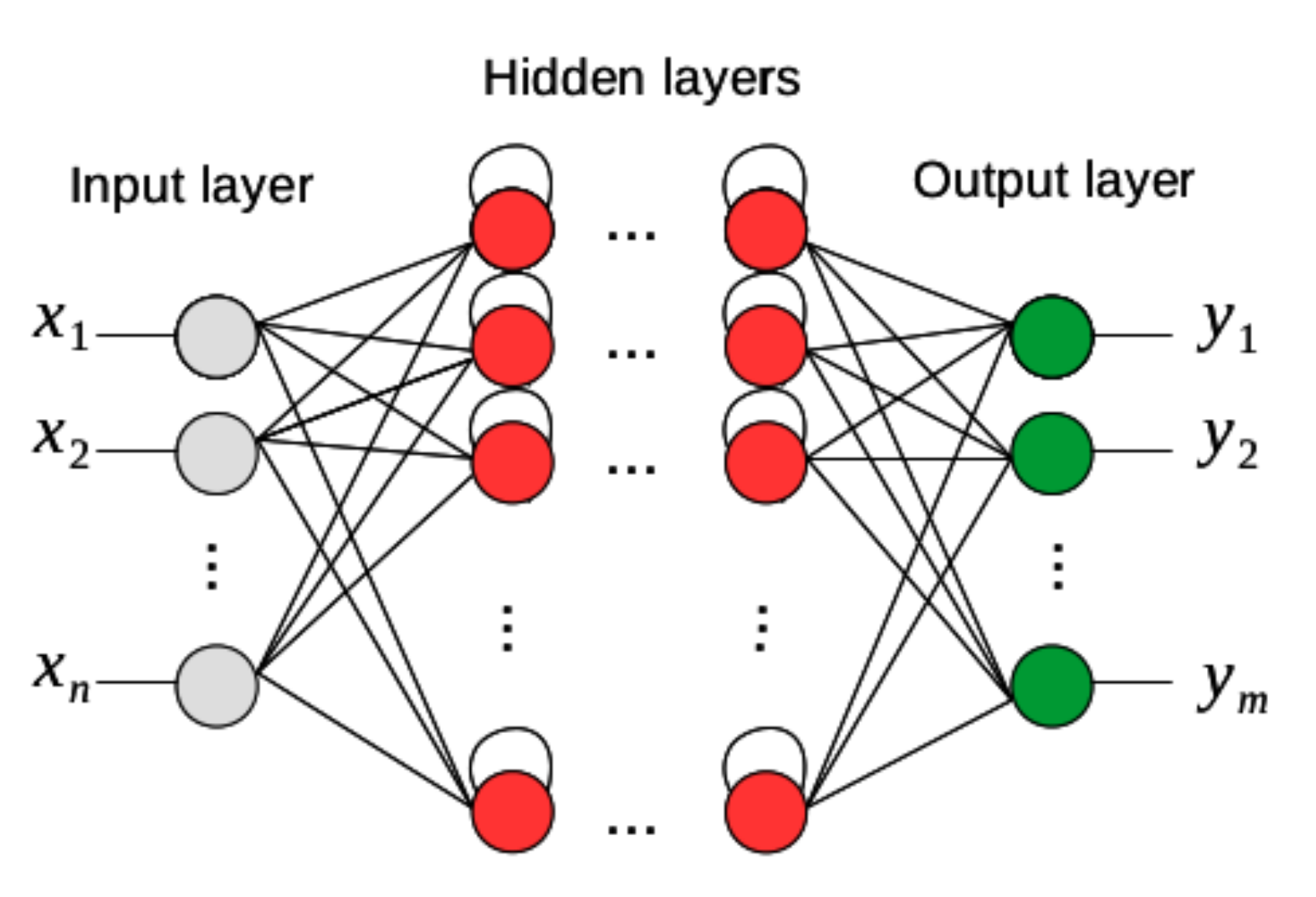}
  \caption{\label{fig:lldppacket} Deep Recurrent Neural Network}\label{drnn}
\endminipage\hfill
\minipage{0.32\textwidth}%
  \includegraphics[scale=0.32]{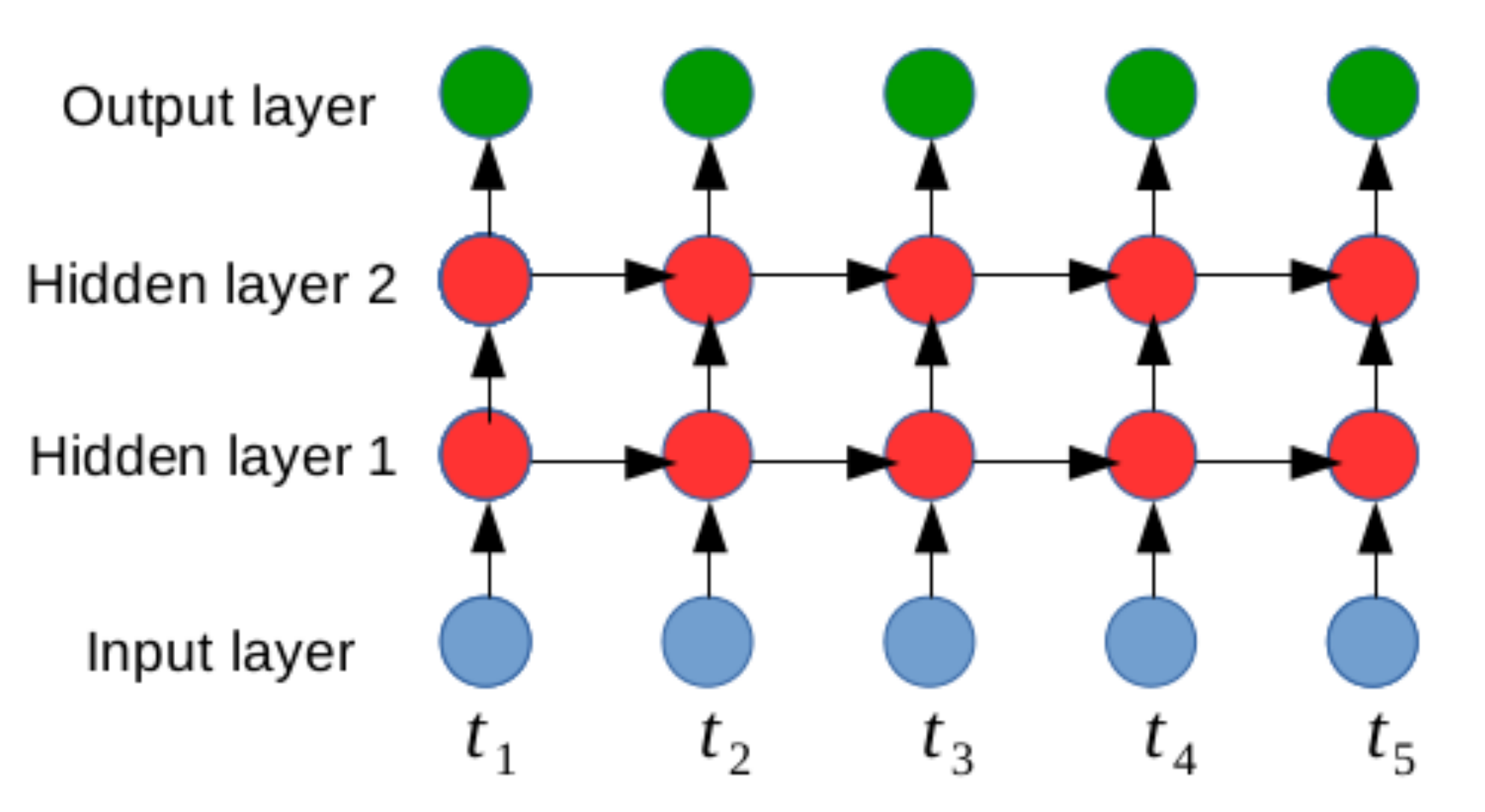}
  \caption{\label{fig:lldppacket} DRNN learning over time}\label{drnnovertime}
  
\endminipage
 \vspace{-1.5em}
\end{figure*}

FFNNs can provide only limited temporal modeling by 
operating on a fixed-size window of TM sequence. 
They can only model the data within the window and are unsuited to
handle historical dependencies.
By contrast, recurrent neural networks or 
deep recurrent neural networks (figure \ref{drnn}) contain cycles that feed back
the network activations from a previous time step as inputs to
influence predictions at the current time step (figure \ref{drnnovertime}).
These activations are stored in the internal states of 
the network as temporal contextual information \cite{sakgoogle}.


However, training conventional RNNs with the gradient-based
back-propagation through time (BPTT) technique is difficult due to
the vanishing gradient and exploding gradient problems. 
The  influence  of  a  given  input  on  the  hidden  layers,  and  therefore
on the network output, either decays or blows up exponentially when cycling
around  the  network's  recurrent  connections. 
These problems limit the capability of RNNs to model the long
range context dependencies to 5-10 discrete time steps between 
relevant input signals and output \cite{sakgooglearxiv}.

To address these problems, an elegant RNN architecture named Long Short-Term Memory
(LSTM) has been designed \cite{lstmoriginalpaper}. 
LSTMs and conventional RNNs have been successfully applied
to sequence prediction and sequence labeling tasks.  LSTM models
have been shown to perform better than conventional RNNs on learning 
context-free and context-sensitive languages for example \cite{language}.

\subsection{LSTM Architecture}
\begin{figure}[h] 
\centering
   \includegraphics[scale=0.295]{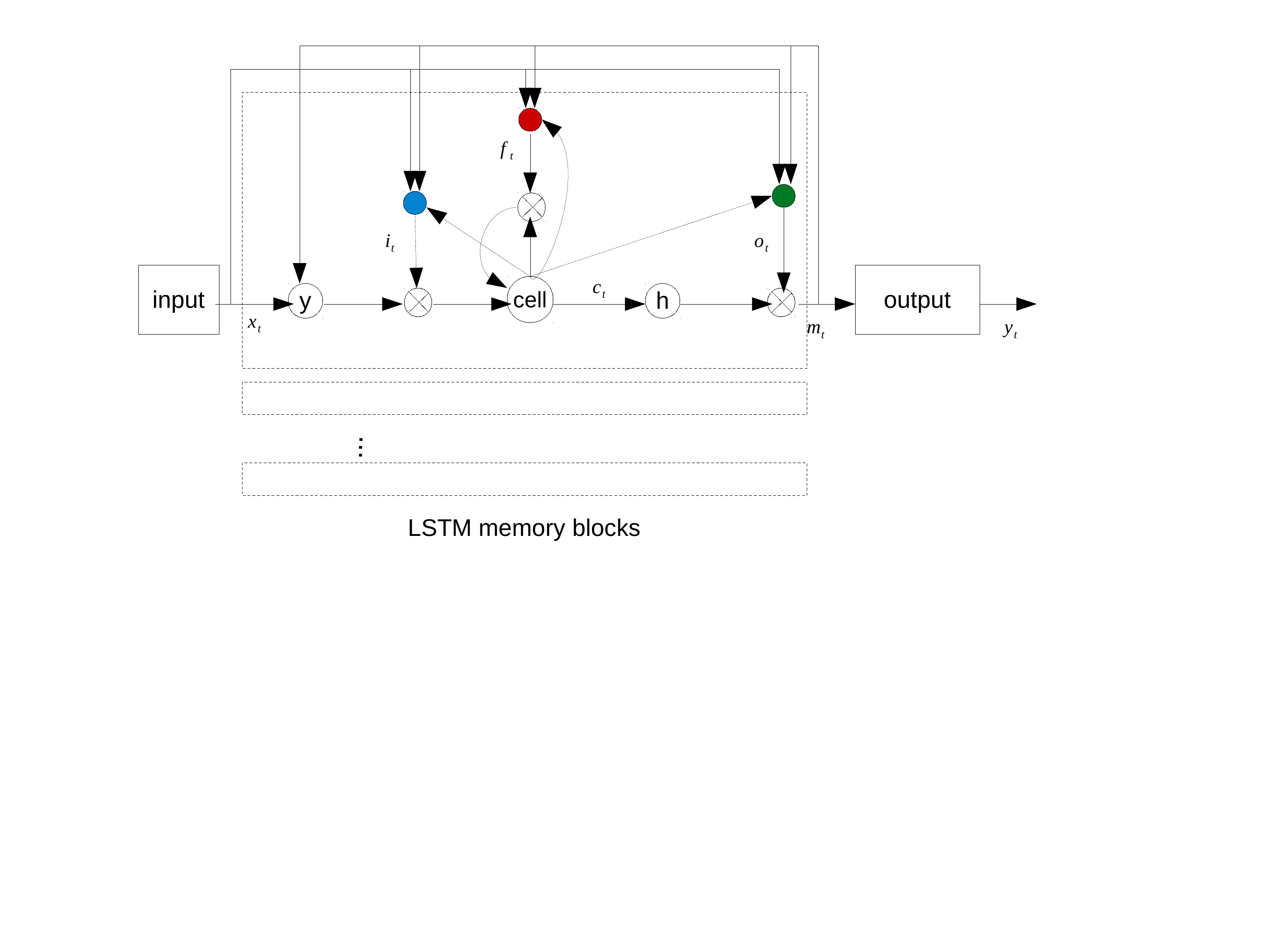}

   \caption{\label{fig:lldppacket} LSTM architecture} \label{archinode}
   \vspace{-1em}
\end{figure}

An LSTM RNN is composed of units 
called memory blocks.  
Each memory block contains memory
cells with self-connections storing (remembering) the temporal state
of the network in addition to special multiplicative units called gates
to control the flow of information.  
Each memory block contains an input gate to control 
the flow of input activations into the memory cell, an 
output gate to control the output flow of cell
activations into the rest of the network and 
a forget gate (figure \ref{archinode}). 

The forget gate scales the internal state of the cell before adding it back to the cell as input through
self recurrent connection, therefore adaptively forgetting
or resetting the cell's memory. The modern LSTM architecture
also contains peephole connections from its internal cells to the gates in
the same cell to learn precise timing of the outputs \cite{peepholes}.

\subsection{LSTM Equations}




An LSTM network maps an input
sequence $x = (x_1 , ..., x_T)$ to an output sequence 
$y =(y_1, ..., y_T)$ by computing the network unit activations using
the following equations iteratively from t = 1 to T.

\begin{equation} \label{eqb}
     i_t = \sigma(W_{ix}x_t + W_{im}m_{t-1} + W_{ic}c_{t-1} + b_i)
\end{equation}
\begin{equation} \label{eqb}
     f_t = \sigma(W_{fx}x_t + W_{fm}m_{t-1}+W_{fc}c_{t-1} + b_f)
\end{equation}
\begin{equation} \label{eqb}
     c_t = f_t  \odot c_{t-1} + i_t \odot g(W_{cx}x_t + W_{cm}m_{t-1} +b_c)
\end{equation}
\begin{equation} \label{eqb}
     o_t = \sigma(W_{ox}x_t + W_{om}m_{t-1} + W_{oc}c_t + b_o)
\end{equation}
\begin{equation} \label{eqb}
     m_t = o_t \odot h(c_t)
\end{equation}
\begin{equation} \label{eqb}
     y_t = \varphi(W_{ym} m_t + b_y)
\end{equation}

Where i, f, o and c are respectively
the input gate, forget gate, output gate and cell activation vectors.
m is the output activation vector. $\odot$ is the element-wise product of the vectors.
g and h are the cell input and cell output activation functions.
tanh and $\varphi$ are the network output activation function.
The b terms denote bias vectors and the W terms denote weight matrices.
and $\sigma$ is the logistic sigmoid function \cite{sakgoogle}.

\section{Traffic Matrix Prediction Using LSTM RNN} \label{tmpredictionusinglstm} 

%

We train a deep LSTM architecture with a 
deep learning method (backpropagation through time algorithm) to learn the traffic characteristics from 
historical traffic data and predict 
the future TM.

\subsection{Problem Statement} \label{problemstatement}
Let N be the number of nodes in the network. The $N$-by-$N$ traffic matrix is denoted by Y 
such as an entry $y_{ij}$ represents the traffic volume flowing from node i to node j. 
We add the time dimension to obtain a structure of N-by-N-by-T tensor (vector of matrices) S such as an entry 
$s_{ij}^t$ represents the volume of traffic flowing from node i to node j at time t,
and T is the total number of time-slots.
The traffic matrix prediction problem is
defined as solving the predictor of $Y^t$ (denoted by $\widehat{Y}^t$) via a series of
historical and measured traffic data set ($Y^{t-1}$, $Y^{t-2}$, $Y^{t-3}$, ..., $Y^{t-T}$).
The main challenge here is how to model the inherent 
relationships among the traffic data set so that one can exactly predict $Y^{t}$.

\subsection{Feeding The LSTM RNN} \label{feeding}
\begin{figure*}[h]
\minipage{0.32\textwidth}
  \includegraphics[scale=0.27]{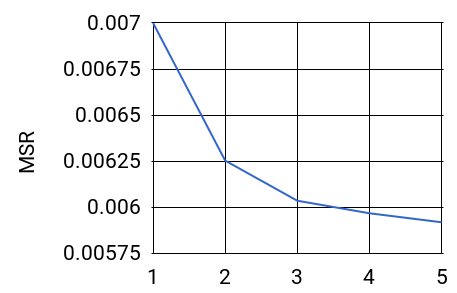}
  \caption{\label{fig:lldppacket} MSE over number of hidden layers (500 nodes each)}\label{msrbynhiddenlayers}
\endminipage\hfill
\minipage{0.32\textwidth}
  \includegraphics[scale=0.27]{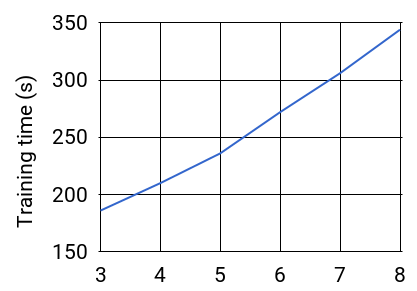}
  \caption{\label{fig:lldppacket} Training time over network depth (20 epochs)}\label{mstbynhidden}
\endminipage\hfill
\minipage{0.32\textwidth}%
  \includegraphics[scale=0.23]{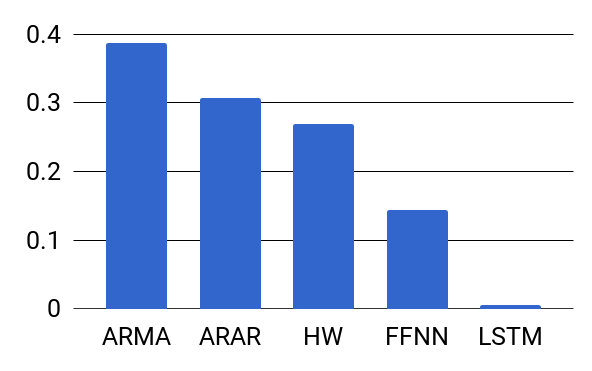}
  \caption{\label{fig:lldppacket} Comparison of prediction methods}\label{compare}
\endminipage
\vspace{-2em}
\end{figure*}

To effectively feed the LSTM RNN, we transform each matrix $Y^t$  to a vector $X^t$ (of size $N\times N$) 
by concatenating its N rows from top to bottom. $X^t$ is called traffic vector (TV).
Note that $x_n$ entries can be mapped to the original $y_{ij}$ using the relation $n=i\times N+j$.
Now the traffic matrix prediction problem is
defined as solving the predictor of $X^t$ (denoted by $\widehat{X}^t$) via a series of
historical measured traffic vectors ($X^{t-1}$, $X^{t-2}$, $X^{t-3}$, ..., $X^{t-T}$).

One possible way to predict the traffic vector $X^t$ is to predict one component $x_n^t$ at a time 
by feeding the LSTM RNN one vector ($x_0^t, x_1^t, ..., x_{N^2}^t)$ at a time.
This is based on the assumption that each OD traffic is independent from all other ODs 
which was shown to be wrong by \cite{nongaussian}. Hence, considering the previous traffic of all 
ODs is necessary to obtain a more correct and accurate prediction of the traffic vector.
\textbf{Continuous Prediction Over Time}:
\begin{figure}[h] \label{window}
\centering
   \includegraphics[scale=0.33]{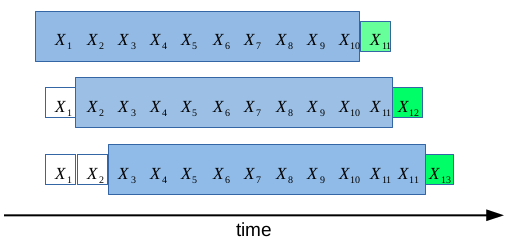}
   \caption{\label{fig:lldppacket} Sliding learning window}\label{window}
    \vspace{-1em}
\end{figure} 
Real-time prediction of traffic matrix requires continuous feeding and learning. 
Over time, the total number of time-slots become too big resulting in high computational complexity.
To cope with this problem, we introduce the notion of learning window (denoted by $W$) 
which indicates a fixed number of previous
time-slots to learn from in order to predict the current traffic vector $X^t$ (Fig. \ref{window}).
We construct the $W$-by-$N^2$ traffic-over-time matrix (that we denote by $M$) by putting together $W$ vectors 
($X^{t-1}$, $X^{t-2}$, $X^{t-3}$, ..., $X^{t-W}$) ordered in time. 
Note that $T\geq W$ ($T$ being the total number of historical matrices) and the number of matrices $M$ is equal to $T/W$.

\subsection{Performance Metric}

To quantitatively assess the overall performance of our LSTM model, 
Mean Square Error (MSE) is used to estimate the prediction accuracy.
MSE is a scale dependent metric
which quantifies the difference between the forecasted
values and the actual values of the quantity being 
predicted by computing the average sum of squared errors:

\begin{equation} \label{phi}
MSE=\frac{1}{N} \sum_{i=1}^N (y_i- \widehat{y}_i )^2
\end{equation}
where $y_i$ is the observed value, $\widehat{y}_i$ is the predicted value
and N represents the total number of predictions.

\section{Experiments and Evaluation} \label{experiments}

We implemented NeuTM as a traffic matrix prediction application on
top of POX controller \cite{pox}. NeuTM's LSTM model is
implemented using Keras library \cite{keras} on top of Google’s TensorFlow machine learning framework \cite{tensorflow}. 
We evaluate the prediction accuracy 
of our method using real traffic data from the G\'EANT 
backbone networks \cite{geant} made up of 23 peer nodes 
interconnected using 38 links (as of 2004).
2004-timeslot traffic matrix data is sampled from the G\'EANT network by 15-min
interval \cite{geantdata} for several months. 

To evaluate our method on short term traffic matrix prediction, 
we consider a set of 309 traffic
matrices.
As detailed in section \ref{feeding}, we transform the matrices to vectors of size $529$ each and we concatenate 
the vectors to obtain the traffic-over-time matrix $M$ of size $309\times 529$. We split $M$ into
two matrices, training matrix $M_{train}$ and validation matrix $M_{test}$ of sizes $263\times 529$ and $46\times 529$ consecutively. 
$M_{train}$ is used to train the LSTM model and $M_{test}$ is used to evaluate and validate its accuracy. Finally, 
We normalize the data by dividing by the maximum value.


Figure \ref{msrbynhiddenlayers} depicts the MSE obtained over different 
numbers of hidden layers (depths). The prediction accuracy is better with deeper networks.
Figure \ref{mstbynhidden} depicts the variation of the training time over different depths. 
Note that it takes less than 5 minutes to train a 6 layers network on 20 epochs.
Finally, figure \ref{compare} compares the prediction error of 
the different prediction methods presented in this paper and shows the superiority of LSTM.

\section{Related Work} \label{relatedwork}
Various methods have been proposed to predict traffic matrix. 
\cite{evaluation} evaluates and compares traditional linear prediction models (ARMA, ARAR, HW)
and neural network based prediction with multi-resolution learning. 
The results show that NNs outperform
traditional linear prediction methods
which cannot meet the accuracy requirements.
\cite{nongaussian} proposes a FARIMA predictor 
based on an $\alpha$-stable non-Gaussian self-similar traffic model. 
\cite{predictionandcorrection} compares three prediction methods: 
Independent Node Prediction (INP), 
Total Matrix Prediction with Key Element Correction (TMP-KEC)
and Principle Component Prediction with Fluctuation Component Correction (PCP-FCC). 
INP method does
not consider the correlations among the nodes, resulting in
unsatisfying prediction error. TMP-KEC method reduces the
forecasting error of key elements as well as that of the total
matrix. PCP-FCC method improves the overall prediction error
for most of the OD flows.

\section{Conclusion}\label{conclusion}

In this work, we have shown that LSTM architectures are well suited for traffic matrix prediction.
We have proposed a data pre-processing and RNN feeding technique that achieves high prediction accuracy 
in a very short training time.
The results of our evaluations show that LSTMs outperforms
traditional linear methods and feed forward neural networks by many orders of magnitude.

\end{document}